\documentclass[aps,prl,twocolumn,superscriptaddress,showpacs]{revtex4-1}

\usepackage[colorlinks,linkcolor=blue,citecolor=red,urlcolor=blue,breaklinks=true]{hyperref}
\usepackage{amsmath,amssymb,graphics,epsfig,epstopdf,color,verbatim,braket,tabularx,breakurl}

\begin{document}

\title{An Anderson impurity interacting with the helical edge states in a quantum spin Hall insulator}

\author{Ru Zheng}
\author{Rong-Qiang He}
\email{rqhe@ruc.edu.cn}
\address{Department of Physics, Renmin University of China, Beijing 100872, China}
\author{Zhong-Yi Lu}
\email{zlu@ruc.edu.cn}
\address{Department of Physics, Renmin University of China, Beijing 100872, China}
%\date{\today}

\begin{abstract}
Using the natural orbitals renormalization group (NORG) method, we have investigated the screening of the local spin of an Anderson impurity interacting with the helical edge states in a quantum spin Hall insulator. We find that there is a local spin formed at the impurity site and the local spin is completely screened by electrons in the quantum spin Hall insulator. Meanwhile, the local spin is screened dominantly by a single active natural orbital. We then show that the Kondo screening mechanism becomes transparent and simple in the
framework of natural orbitals formalism. We project the active natural orbital respectively into real space and momentum space to characterize its structure. And we confirm the spin-momentum locking property of the edge states based on the occupancy of a Bloch state in the edge to which the impurity couples. Furthermore, we study the dynamical property of the active natural orbital represented by the local density of states, from which we observe the Kondo resonance peak.
\end{abstract}


\maketitle

The quantum spin Hall (QSH) insulator, first discovered in HgTe/CdTe quantum wells\cite{Konig2007} following its theoretical prediction\cite{Bernevig2006}, has been tremendously investigated. It is characterized by a full insulating gap in the bulk but one-dimensional gapless conducting edge states with opposite spins counterpropagating at each edge, where the spin-orbit coupling (SOC) plays an essential role\cite{Hasan2010,Qi2011}. The edge states are of nontrivial helical liquid\cite{CWu2006} with the quantized conductance of $G = e^2/h$ for each single helical edge state protected by the time-reversal symmetry (TRS).

Since the TRS protects the helical edge states from backscattering, they are robust against weak interactions and perturbations that preserve the TRS\cite{Kane20052,CWu2006,Xu2006}. Particularly, a single edge electron cannot be backscattered by an impurity without internal degrees of freedom. However, a quantum impurity interacting with the helical edge states makes a nontrivial perturbation which allows backscattering accompanied by a spin-flip scattering. The effect of a quantum impurity on the conductance of the helical edge states has been investigated\cite{CWu2006,Maciejko2009,Yoichi2011,Joseph2012,Erik2013}. Maciejko and coworkers\cite{Maciejko2009} argued that the conductance of a helical edge state exhibits a logarithmic temperature dependence at high temperature resulting from the backscattering by a magnetic impurity, while it restores to the quantized value $e^2/h$ at $T = 0$ due to the formation of a Kondo singlet with a complete screening of the impurity spin by the helical liquid. On the other hand, the influence of the SOC on the Kondo effect in QSH insulators has also been studied\cite{Zitko2011, Zarea2012,Isaev2012,Kikoin2012,Grap2012,Mastrogiuseppe2014,Wong2016,Sousa2016}. Nevertheless, it lacks that works focus on the screening of the local spin by the Kondo cloud for a quantum impurity interacting with a helical liquid in a quantum spin Hall insulator.

The Kane-Mele (KM) model\cite{Kane20052,Kane20051} is a spinful model preserving the TRS and exhibits a QSH effect. The ground state of the KM model defined on a graphene ribbon is a QSH insulator with the helical edge states in each edge. This model is related to the spinless Haldane model\cite{Haldane1988} which breaks the TRS and it can be considered as two copies of the Haldane model. A single Anderson impurity coupled to a zigzag graphene edge at a finite temperature has been investigated by the Quantum Monte Carlo simulations method\cite{Assaad2013,Hu2013}. On the other hand, the spatial structure of spin correlations around an Anderson impurity at the edge of a silicene-like topological insulator has been studied by the density matrix renormalization group method\cite{Andrew2017b}. In this paper, using the newly developed natural orbitals renormalization group method\cite{He2014}, we study the mechanism underlying the screening of the local spin of an Anderson impurity interacting with the helical edge states represented by the KM model at zero temperature.

\begin{figure}[h!]
\centering
\includegraphics[width=0.8\columnwidth]{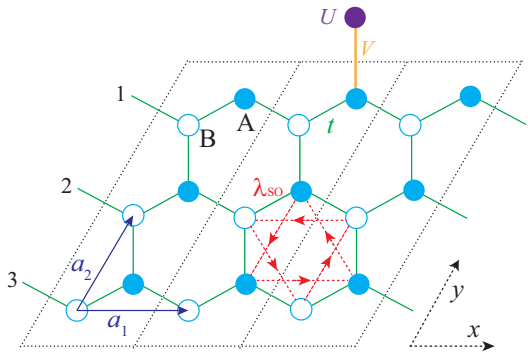}
\caption{\label{fig:KMAM}(color online) Sketch of a KM model with an Anderson impurity at an edge in a honeycomb lattice ribbon. The ribbon is periodical along the $x$ direction and open along the $y$ direction, i.e., the edges are of zigzag. The unit cell of honeycomb lattice with zigzag edge is shown as the dashed black parallelogram. The cyan filled circles denote sublattice A and open circles sublattice B of the honeycomb lattice. The green lines denote the nearest-neighbor hopping $t$ connecting the two sublattices. The spin-orbital coupling term connecting the same sublattice is denoted as the red dashed arrows, and its sign is associated with $\nu_{ij}$ in Eq.~\ref{eq:KM}. $a_1$ and $a_2$ represent the unit cell vectors of the honeycomb lattice. The Anderson impurity is marked by the filled purple circle and it hybridizes with a site belonging to sublattice A along the edge with hybridization $V$, the onsite Hubbard $U$ is at the impurity site.}
\end{figure}

{\em Model and Method.}---The total Hamiltonian of the KM model with an Anderson impurity at an edge is given by
\begin{equation}
H = H_{\text{KM}} + H_{\text{imp}} + H_{\text{hyb}},
\label{eq:Model}
\end{equation}
where $H_{\text{KM}}$ denotes the KM model defined in a honeycomb lattice ribbon with zigzag edges, $H_{\text{imp}}$ denotes an Anderson impurity nearby an edge, $H_{\text{hyb}}$ represents the hybridization between the Anderson impurity and a site at the edge coupled with the impurity. For simplicity, we consider the case where the impurity is coupled to an edge of sublattice A, as shown in Fig.~\ref{fig:KMAM}.

The Hamiltonian of the KM model $H_{\text{KM}}$ includes two parts as follows,
\begin{equation}
\begin{array}{l}
H_{\text{KM}} = H_t + H_{\text{SO}}, \\
H_t = -t\sum\limits_{\langle ij\rangle\sigma}c_{i\sigma }^ {\dagger}{c_{j\sigma }} ,\\
H_{\text{SO}} = i\lambda_{\text{SO}}\sum\limits_{\langle\langle ij\rangle\rangle \alpha\beta}\nu_{ij}c_{i,\alpha}^{\dagger}\sigma_{\alpha\beta}^zc_{j,\beta}.
\end{array}
\label{eq:KM}
\end{equation}
Here $H_t$ describes the tight-binding band in a honeycomb lattice, $H_{\text{SO}}$ denotes the spin-orbital coupling part. $c_{i\sigma}^{\dagger}$ creates an electron at site $i$ with spin component $\sigma = \uparrow, \downarrow$ and $t$ is the nearest-neighbor hopping parameter with $t$ set to 1 in all the following calculations. $\langle\langle i,j\rangle\rangle$ denotes the next-nearest-neighbor hopping with a complex hopping integral and $\lambda_{\text{SO}}$ represents the strength of spin-orbital coupling. The parameter $\nu_{ij} = -\nu_{ji} = \pm 1$ depends on the orientation of the two nearest-neighbor bonds that an electron hops from site $j$ to $i$, namely $\nu_{ij} = +1$ if the electron turns left in the hopping from site $j$ to $i$ and $\nu_{ij} = -1$ if it turns right, as shown in Fig.~\ref{fig:KMAM}. In the spin-orbital coupling part $H_{\text{SO}}$, $\sigma_{\alpha\beta}^z$ is the $z$ Pauli matrix which further distinguishes the spin up and spin down states with opposite next-nearest-neighbor hopping amplitude. The  ground state of the KM model on a half-filled honeycomb lattice ribbon with zigzag edges describes a quantum spin Hall insulator.

The Hamiltonians of an Anderson impurity and its hybridization with edge electrons are given by
\begin{equation}
\begin{array}{l}
H_{\text{imp}} = -\mu(n_{\text{imp},\uparrow} + n_{\text{imp},\downarrow}) + Un_{\text{imp},\uparrow}n_{\text{imp},\downarrow},\\
H_{\text{hyb}} = V\sum\limits_\sigma(c_{r_i,\sigma}^{\dagger}c_{\text{imp},\sigma} + c_{\text{imp}, \sigma}^{\dagger}c_{r_i,\sigma}).
\end{array}
\label{eq:Anderson impurity}
\end{equation}
Here $n_{\text{imp},\uparrow}(n_{\text{imp},\downarrow})$ represents the impurity occupancy number operator with spin component $\sigma=\uparrow(\downarrow)$, $-\mu$ denotes the single-particle energy of the impurity, $U$ denotes the strength of Hubbard interaction at the impurity site. The impurity hybridizes with a site located at $r_i$ in the edge with hybridization $V$. $c_{r_i,\sigma}^ {\dagger}(c_{r_i,\sigma})$ and $c_{\text{imp},\sigma }^ {\dagger}(c_{\text{imp},\sigma})$ represent the creation (annihilation) operators at the site hybridized with the impurity and at the impurity site respectively with spin component $\sigma=\uparrow,\downarrow$.

The total Hamiltonian $H$ preserves both charge ${\rm U}(1)_{\rm charge}$ symmetry and spin ${\rm U}(1)_{\rm spin}$ symmetry, even though the spin-rotation symmetry ${\rm SU}(2)$ is broken by the spin-orbital coupling term. Thus, the $z$-component of the total spin is conserved. Furthermore, the whole system preserves the TRS. In the calculations, we considered a hall-filling case for the band structure with the spin-orbital coupling $\lambda_{\rm SO}=0.1$. A particle-hole symmetric representation of the Hubbard interaction that sets $\mu = U/2$ for the half-filling case was employed. We set the hybridization $V=0.5$.

We employed a newly developed numerical many-body approach, namely the natural orbitals renormalization group (NORG), which works in the Hilbert space constructed from a set of natural orbitals. In practice, the realization of the NORG essentially involves a representation transformation from site representation into natural orbitals representation through iterative orbital rotations (see Ref.~\onlinecite{He2014} for details). The NORG is naturally appropriate for dealing with a quantum impurity system. As shown in Ref.~\onlinecite{He2014}, for a quantum impurity system most of the natural orbitals exponentially rush into doubly occupancy or empty, while only a small number of natural orbitals, equal to the number of impurities, deviate well from full occupancy and empty, namely active natural orbitals with half filling or nearly, which play a substantial role in constructing the ground state wave function.

\begin{figure}[htp!]
\centering
\includegraphics[width=1\columnwidth]{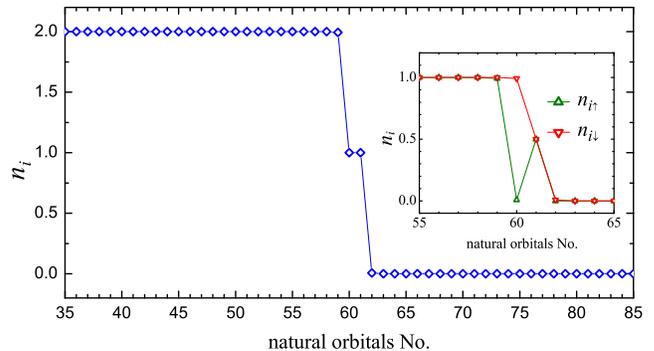}
\caption{\label{fig:NOOcc}(color online) Natural orbitals occupancy distribution $n_i$ for the ground state with the Hubbard $U=4.0$. The inset shows the corresponding occupancies for spin up $n_{\uparrow}$ and spin down $n_{\downarrow}$. The calculation was carried out in an $S^z_{\rm tot}=-1/2$ subspace and the honeycomb lattice ribbon size we employed is $L = L_x \times L_y$ with $L_x=20$ and $L_y=3$.}
\end{figure}

{\em Results.}---Figure~\ref{fig:NOOcc} shows the calculated natural orbitals occupancies $n_i$ for the ground state carried out in an $S^z_{\rm tot}=-1/2$ subspace (here the number of electrons is odd). As we see, all the natural orbitals exponentially rush into full occupancy or empty except for two natural orbitals, for which we find that the occupancies for one natural orbital are nearly 0 for spin up and 1.0 for spin down, i.e., $n_{\uparrow}\approx 0$ and $n_{\downarrow}\approx 1.0$, while the occupancies are both 0.5 for the other natural orbital, i.e., $n_{\uparrow}=0.5$ and $n_{\downarrow}=0.5$. Therefore, only the natural orbital with occupancies $n_{\uparrow} = 0.5$ and $n_{\downarrow}= 0.5$ is active, indexed as the 61-st natural orbital in Fig.~\ref{fig:NOOcc}. Thus, there is only one active natural orbital for the edge states interacting with an magnetic impurity. This makes the Kondo screening mechanism transparent and simple, as presented in the following.

Due to the spin-orbital coupling, we consider the screening of the $z$-component of the impurity spin, which is determined by the spatially extended Kondo screening cloud. Specifically, this is described by the spin correlation between the magnetic impurity spin and the electron spin. Here we use ${\bf{S}}_{\text{ANo}}$ to denote the spin of electrons occupying the active natural orbital, and ${\bf{s}}_i$ the spin of electrons occupying the $i$-th site. Accordingly, we calculated the spin correlation $\langle {\bf{S}}_{\text{imp}}^z{\bf{s}}_{\text{ANo}}^z\rangle$ between the active natural orbital and magnetic impurity spin as well as the following integrated spin correlation
\begin{equation}
\langle{\bf{S}}_{\text{imp}}^z{\bf{s}}^z \rangle= \sum\limits_{i=1}^L \langle {\bf{S}}_{\text{imp}}^z{\bf{s}}_i^z\rangle= \sum\limits_{m=1}^L \langle {\bf{S}}_{\text{imp}}^z{\bf{s}}_m^z\rangle,
\label{eq:integrated-screening}
\end{equation}
where ${\bf{s}}_m$ denotes the spin of electrons occupying the $m$-th natural orbital with $L=L_x\times L_y$ being the system size. On the other hand, the local spin $s$ of electrons at the impurity site is determined by
\begin{equation}
\langle({{\bf{S}}_{\text{imp}}^z})^2\rangle = \frac{s(s+1)}{3}.
\label{eq:integrated-screening}
\end{equation}

\begin{figure}[htp!]
\centering
\includegraphics[width=1\columnwidth]{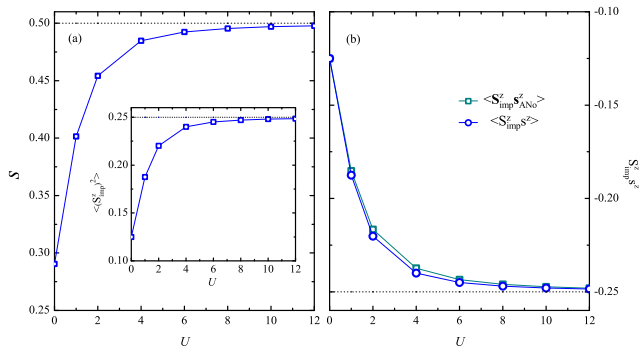}
\caption{\label{fig:ScreeningMechanism}(color online) (a) Local spin $s$ of electrons at the impurity site and (b) spin correlation $\langle {\bf{S}}_{\text{imp}}^z{\bf{s}}_{\text{ANo}}^z\rangle$ between the active natural orbital and magnetic impurity as well as the integrated spin correlation $\langle {\bf{S}}_{\text{imp}}^z{\bf{s}}^z\rangle$ as functions of the Hubbard $U$. The inset in (a) shows the $z$-component of the local moment $\langle ({{\bf{S}}_{\text{imp}}^z})^2 \rangle$ at the impurity site. The dotted lines mark corresponding values in the large Hubbard $U \to \infty$ limit. The honeycomb lattice ribbon size we employed is $L = L_x\times L_y$ with $L_x=50$ and $L_y=3$.}
\end{figure}

As we see from Fig.~\ref{fig:ScreeningMechanism}, the local spin $s$ is formed at the impurity site and it increases with $U$. In the large Hubbard $U$ limit $U\to \infty$, the local spin $s$ goes to $\frac{1}{2}$ with double occupancy suppressed. Meanwhile the calculations show that $\langle{\bf{S}}_{\text{imp}}^z\rangle=0$ in the $S^z_{\rm tot}=-1/2$ subspace. This means that no free local moment on the impurity site can be polarized and hence the impurity spin is completely screened by electrons in the graphene ribbon. In such a case, no spin-flip scattering exists for an electron propagating to the impurity from far distance, the conductance of a helical edge state will be thus quantized as $e^2/h$. The thermal fluctuations at a finite temperature will partially destruct the screening and make the conductance decreasing from $e^2/h$. Moreover, the impurity is in antiferromagnetic coupling with the electrons occupying the active natural orbital. And by comparison, the spin correlation $\langle {\bf{S}}_{\text{imp}}^z{\bf{s}}_{\text{ANo}}^z\rangle$ nearly coincides with the integrated spin correlation $\langle{\bf{S}}_{\text{imp}}^z{\bf{s}}^z\rangle$, indicating that the active natural orbital dominantly screens the impurity spin in $z$ direction. In the large Hubbard $U$ limit $U\to \infty$, the spin correlation $\langle {\bf{S}}_{\text{imp}}^z{\bf{s}}_{\text{ANo}}^z\rangle \to -1/4$ as well as the integrated spin correlation $\langle {\bf{S}}_{\text{imp}}^z{\bf{s}}^z\rangle \to -1/4$, which demonstrates that the active natural orbital screens the impurity spin solely.

To characterize the structure of the active natural orbital, we project it into real space (site representation, namely Wannier representation) and momentum space, respectively. To do so, we consider the following Fourier transformation of the electron operator in a graphene ribbon,
\begin{equation}
\begin{array}{l}
c_{nm,\alpha,A}^{\dagger}=\frac{1}{\sqrt L_x}\sum\limits_k e^{-ikX_{nm,A}}a_{n,\alpha}^{\dagger}(k), \\
c_{nm,\alpha,B}^{\dagger}=\frac{1}{\sqrt L_x}\sum\limits_k e^{-ikX_{nm,B}}b_{n,\alpha}^{\dagger}(k).
\end{array}
\label{eq:Fourier}
\end{equation}
Noting that the graphene ribbon with zigzag edges is periodical in the $x$ direction (Fig.~\ref{fig:KMAM}). Here $(nm,A)$ or $(nm,B)$ labels a lattice site, $(nm)$ labels a unit cell with A or B labelling sublattice A or B, $a_{n,\alpha}^{\dagger}$ or $b_{n,\alpha}^{\dagger}$ creates an electron with spin component $\alpha = \uparrow, \downarrow$ at sublattice A or B.

Thus, we can easily obtain the amplitude $|w_i^m|^2$ of the $m$-th natural orbital projected into the $i$-th Wannier orbital by $d_m^{\dagger}=\sum_{i=1}^Nw^m_ic_i^{\dagger}$ from the NORG transformation. Likewise, we can project the $m$-th natural orbital into a Bloch state with wavevector $k$ in momentum space via $d_m^ {\dagger}=\sum_{k}u_k^mc_k^{\dagger}$, where $c_k^{\dagger}$ denotes the creation operator in momentum space. The $u_k^m$ can be obtained from the NORG transformation in combination with the Fourier transformation shown in Eq.~(\ref{eq:Fourier}) for $c_k^{\dagger}$. Meanwhile, the occupancy $n(k)$ of a Bloch state with wavevector $k$ can be obtained by $n(k)=\sum_{m=1}^Nn_{m}|u_k^m|^2$ with $n_{m}$ being the occupancy number of the $m$-th natural orbital.

\begin{figure}[htp!]
\centering
\includegraphics[width=1\columnwidth]{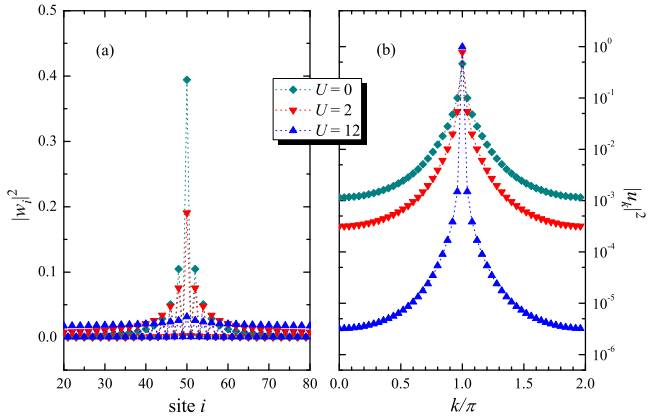}
\caption{\label{fig:Amplitude}(color online) Amplitude of the active natural orbital projected into (a) real space and (b) momentum space in the upper edge with the Hubbard $U=0$, 2, and 12 respectively. The Fermi energy of the KM model is $\varepsilon_{k_F}=0$ with the Fermi wave vector $k_F=\pi$. The honeycomb lattice ribbon size we employed is $L = L_x\times L_y$ with $L_x=50$ and $L_y=3$.}
\end{figure}

We now consider the sum weight $W_{A(B)}$ contributed by the sites belonging to sublattice A(B) in the edge to which the magnetic impurity couples (the upper edge), namely $W_{A(B)}=\sum_i|w_i^{\rm ANo}|^2$ with site-$i$ belonging to sublattice A(B) located at the upper edge. The calculations show that $W_A \ge 95\%$ and $W_B \ge 2\%$ in all cases. This means that the impurity spin is screened mainly by the sites or Wannier orbitals belonging to sublattice A in the upper edge and that the Kondo screening cloud extends mainly in the upper edge.

We then project the active natural orbital respectively into momentum space and real space in the upper edge. The amplitude of the active natural orbital projected into real space $|w_i|^2$ and momentum space $|u_k|^2$ are shown in Fig.~\ref{fig:Amplitude}(a) and Fig.~\ref{fig:Amplitude}(b), respectively. For the Hubbard $U=0$, the site (labeled as site 50) which links directly with the impurity dominantly constitutes the active natural orbital, indicating that the active natural orbital is very localized. As the Hubbard $U$ increases, all the sites namely Wannier orbitals belonging to sublattice A in the upper edge tend to equally compose the active natural orbital, indicating that the active natural orbital becomes delocalized. In contrast, in momentum space, the single-particle states near the Fermi energy (low-energy excitations) become dominant to participate in constituting the active natural orbital with the Hubbard $U$ increasing. Especially, the single-particle states at the Fermi energy level solely constitute the active natural orbital in the large Hubbard $U$ limit $U \to \infty$, i.e., $|u_{k_F=\pi}|^2 \to 1.0$ when $U \to \infty$.

\begin{figure}[htp!]
\centering
\includegraphics[width=1\columnwidth]{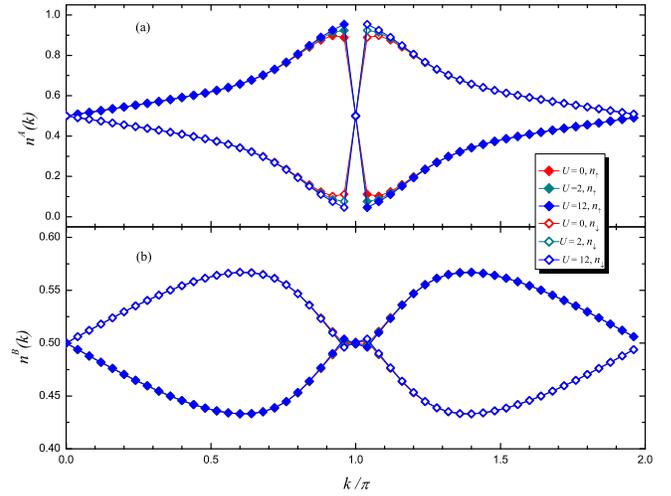}
\caption{\label{fig:EnergyOccupancy}(color online) Occupancy $n(k)$ of a Bloch state with wavevector $k$ for (a) sublattice A and (b) sublattice B in the upper edge with the Hubbard $U=0$, 2, and 12 respectively. The Fermi energy of the KM model is $\varepsilon_{k_F}=0$ with the Fermi wave vector $k_F=\pi$. The honeycomb lattice ribbon size we employed is $L = L_x\times L_y$ with $L_x=50$ and $L_y=3$.}
\end{figure}

Results of the occupancy $n(k)$ of a Bloch state with wavevector $k$ in the upper edge are shown in Fig.~\ref{fig:EnergyOccupancy}, where $n^{\rm A}(k)$ and $n^{\rm B}(k)$ denote the occupancy for sublattice A and B respectively. As we see, the electrons with different spins are separated, resulting from the spin-orbital coupling. The difference $|n_{\uparrow}(k)-n_{\downarrow}(k)|$ of the occupancies for different spins near the Fermi wave vector $k_F=\pi$ is as large as about 1.0 for sublattice A, while that for sublattice B is about 0.15, in contrast with the occupancy for a Bloch state in the bulk namely $n_{\uparrow}=n_{\downarrow}=0.5$. This also means that the edge states reside mainly in sublattice A. Moreover, the Hubbard $U$ changes the occupancies $n^{\rm A}(k)$ near the Fermi wave vector $k_F=\pi$, while the others remain almost unchanged, demonstrating that the Hubbard $U$ has effect mostly on the edge states. Here we emphasize that the occupancies at the Fermi wave vector $k_F=\pi$, namely the time-reversal invariant point, show $n^{\rm A}_\uparrow(k_F) = n^{\rm A}_\downarrow(k_F)$ and $n^{\rm B}_\uparrow(k_F) = n^{\rm B}_\downarrow(k_F)$, which is due to the Kramers degeneracy. Moreover, the electron spins are opposite with respect to the Fermi wave vector $k_F=\pi$, indicating the spin-momentum locking of the edge states.

In the Lehmann representation, the local one-particle Green's function defined at sites or orbitals can be expressed as
\begin{equation}
\begin{split}
G_{i,\sigma}(\omega) =& \langle0|c_{i\sigma}\frac{1}{\omega+i\eta-(H-E_0)}c^\dagger_{i\sigma}|0\rangle  \\
                       & +\langle0|c^\dagger_{i\sigma}\frac{1}{\omega+i\eta+(H-E_0)}c_{i\sigma}|0\rangle,
\label{eq:Greenfunction}
\end{split}
\end{equation}
where $|0\rangle$ and $E_0$ mean the ground state and ground-state energy respectively. The local density of states (LDOS) is given by $\rho_{i,\sigma}(\omega)=-\frac{1}{\pi}{\text{Im}}(G_{i,\sigma})$ with a Lorentzian broadening factor $\eta \to 0$. We focus on the LDOS $\rho_{{\rm ANo},\uparrow}(\omega)$ at the active natural orbital, and we calculate this quantity using the correction vector method\cite{Kuhner1999,Jeckelmann2002}.

\begin{figure}[htp!]
\centering
\includegraphics[width=0.8\columnwidth]{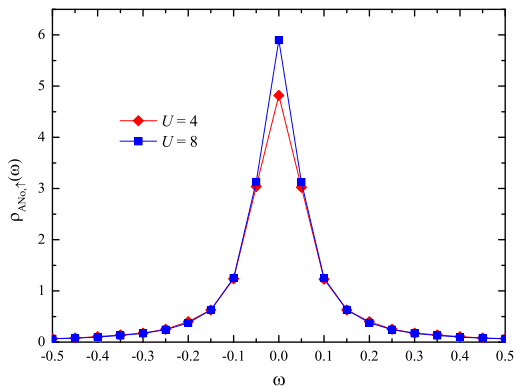}
\caption{\label{fig:LDOSANo}(color online) Local density of states $\rho_{{\rm ANo},\uparrow}(\omega)$ at the active natural orbital for the Hubbard $U=4$ and 8 and with the Lorentzian broadening factor $\eta=0.05$. The Fermi energy of the KM model is $\varepsilon_{k_F}=0$ with the Fermi wave vector $k_F=\pi$. The calculations were carried out on a honeycomb lattice ribbon of size $L = L_x\times L_y$ with $L_x=20$ and $L_y=3$.}
\end{figure}

Figure~\ref{fig:LDOSANo} shows the LDOS $\rho_{{\rm ANo},\uparrow}(\omega)$ at the active natural orbital for the Hubbard $U=4$ and 8 with $\eta=0.05$. From Fig.~\ref{fig:LDOSANo}, we see a peak at $\omega=0$, i.e., at the Fermi energy. This is actually the Kondo resonance peak, and the resonance peak is enhanced with increasing the Hubbard $U$.

{\em Conclusion.}---We have investigated an Anderson impurity interacting with the helical edge states in a quantum spin Hall insulator using the NORG method. We find that there is a local spin formed at the impurity site and then the impurity spin is completely screened by electrons in the quantum spin Hall insulator, namely the Kondo screening effect. Meanwhile, there exists a single active natural orbital which dominantly screens the local spin. And the active natural orbital is well characterized by projecting it respectively into real space and momentum space. This makes the Kondo screening mechanism transparent and simple. We further confirm the spin-momentum locking property of the edge states. And the Kondo resonance peak is also observed from the local density of states at the active natural orbital. Our study is well helpful to further studying magnetic impurities in quantum spin Hall insulators.

\begin{acknowledgments}

This work was supported by National Natural Science Foundation of China (Grants No. 11474356 and No. 11774422). R.Q.H. was supported by the Fundamental Research Funds for the Central Universities, and the Research Funds of Renmin University of China. Computational resources were provided by National Supercomputer Center in Guangzhou with Tianhe-2 Supercomputer and Physical Laboratory of High Performance Computing in RUC.

\end{acknowledgments}

\bibliography{KMAMBib}

\end{document}